\documentstyle[12pt]{article}
\def\beq{\begin{equation}}
\def\eeq{\end{equation}}

\author{{\large M.Yu.Khlopov{\small $^{1,2,3}$}\thanks{e-mail: mkhlopov@orc.ru}
 ,~A.S.Sakharov{\small $^{1,3}$}\thanks{e-mail: sakhas@landau.ac.ru}
 ~and~A.L.Sudarikov{\small $^{1,3}$}\thanks{e-mail: sudar@orc.ru}}}
\title{{\huge{\bf CosmoParticle Physics:
Basic Principles and Prospects for Future Development }}}
\date{{\small{\it $^1$Center for CosmoParticle Physics "Cosmion"\\ 
$^2$Institute of Applied Mathematics,
Miusskaya Pl.4, 125047 Moscow, Russia\\
$^3$Moscow Engineering Physics Institute 
(Technical University),  Kashirskoe Sh.31, 115409 Moscow, Russia}}}

\begin{document}
\maketitle

\begin{abstract}
COSMOPARTICLE PHYSICS is the specific crossdisciplinary field of science, studying  
foundations of particle physics and cosmology in the combination of indirect cosmological, 
astrophysical and physical signatures of their fundamental relationship. The possibilities to 
elaborate unique theoretical grounds for cosmology and particle physics and to study quantitative 
relationships between cosmological and laboratory effects follow from the basic principles of 
cosmoparticle physics and open new interesting fields of scientific research in its future 
development. 
\end{abstract}

\section{Principles of cosmoparticle physics} 

CosmoParticle Physics studies mutual relationship and fundamental physical grounds of Cosmology 
and Particle Physics \cite{1}. It provides unified treatment of the basic laws of the Universe and 
elementary particles, establishes mutual correspondence between them and probes the fundamental 
nature of micro- and macro-worlds in the proper combination of its indirect physical, astrophysical 
and cosmological effects. It offers the nontrivial way out of the wrong circle of problems, to which 
fundamental physics comes in its one-dimensional development.
 
Cosmoparticle physics is now being formed into selfconsistent new science, following internal 
basic principles in its future development. This development revives the tradition of Natural 
philosophy of the universal knowledge, the tradition to consider the world in its universal 
completeness and unity.
 
Cosmoparticle physics reproduces in the largest and smallest scales the general feature of the 
fundamental physics: the mutual correspondence between microscopic and macroscopic 
descriptions, say, between thermodynamics, atomic theory, hydrodynamics and kinetics, or between 
the fundamental macroscopic and microscopic quantities, e.g., between the Avogadro number and 
the mass of proton. However, at the level of fundamental cosmology and particle physics this 
correspondence acquires the new quality of their unity.

That is why the first basic principle of cosmoparticle physics is the idea of a world system, treating 
in the unique framework the foundations of macro- and micro-physics. The second principle 
assumes, that the world system establishes strict quantitatively definite mutual correspondence 
between fundamental cosmological, astrophysical and micro-physical laws, i.e. postulates the 
quantitatively definite correspondence between the structures at macro- and micro- levels. Finally, 
the third principle assumes, that the set of world system parameters does not exceed the number of 
its macro- and micro-scopic signatures.

One may easily find, that the first principle simply postulates the existence of a world system, 
whereas the two other principles specify its necessary properties. The crucial point in this approach 
is multidimensional solution, offered by the cosmoparticle physics to the problems, both cosmology 
and particle theory face on. It may be shown, that this approach naturally embeds all the widely 
known existing trends in studying links between cosmology and particle physics, such as 
astroparticle physics, theories of everything, particle astrophysics, cosmoarcheology.

Here we'd like to specify some new types of links, following with necessity from the basic 
principles of cosmoparticle physics and lying outside these widely discussed trends.

\section{Unified models of cosmology and particle physics}
 
Intensive efforts to construct the finite Theory of Everything, undertaken last decade on the base of 
Superstring models, have not lead, unfortunately, to extensive theoretical framework, putting 
together the modern cosmology and particle physics into the detailed and quantitatively definite 
picture. The point is that the space of classical string vacuum has a vary large degeneracy, and there 
is no objective criteria that distinguishes a particular string vacuum among the numerous 
possibilities. The mathematical complexity is multiplied by the enormous variety of possible 
embeddings of the Standard model (SM) of particle interactions into the structure of superstring 
models. Indeed, the guiding principle of superstring  phenomenology is very simple: it is to 
reproduce the SM within the effective low energy field theory of a string model. Since only general 
features such as the gauge group, number of families, etc. are considered, it leads to numerous 
possibilities for embedding the SM in superstring phenomenology. For example \cite{kaku}, within 
the framework of perturbative heterotic superstring, the total rank of the gauge group (for  $N=1$, 
space-time supersymmetric models) can be as large as $22$. After the SM 
$SU(3)_C\bigotimes SU(2)_W\bigotimes U(1)_Y $
symmetry with the rank $4$ is reproduced, the rank of the residual gauge symmetry can be still as 
large as $18$. Taking into account that the number of models grows (roughly) as a factorial of the  
rank of the residual gauge symmetry, it becomes clear that we need additional arguments to restrict 
the amount of models. One of them is to use grand unification and to embed the SM symmetry 
within a simple gauge group $G\supset SU(3)_C\bigotimes SU(2)_W\bigotimes U(1)_Y$. To break 
the grand unified gauge group $G$ down to that of the SM an adjoint representation of Higgs fields 
must be present in effective field theory among the light degrees of freedom. In perturbative 
heterotic superstring such states in the massless spectrum are compatible with $N=1$ 
supersymmetry and chiral fermions only if the  grand unified gauge group is realized via a current 
algebra at level $k>1$ (see \cite{2}). This condition leads to reduction of the total rank of the gauge 
group, and, therefore, restricts the number of possible models. However, for example, for a grand 
unified gauge group $G=SO(10)$ with, $k=3$, the rank of the residual gauge symmetry can be still 
as large as $7$. Thus even grand unification constraint allows unacceptable amount of SM 
embedding. In the case of more sophisticated and extensive string models the ambiguity grows, 
making virtually impossible to use the main advantage of the string theory -- to calculate all the 
fundamental macro- and microphysical quantities from the first principles.
 
Moreover, however extensive String models are, they do not represent the most general embedding 
for the particle physics and the physics of space-time. The following motivations illustrate some 
idea on the possible form of such a general framework.

Events are basic elements of space-time in relativistic theory. The intervals between them maintain 
the geometry of space-time. So it seems physically meaningful to treat material processes, causing 
the events, together with the space-time, they take place in. But such mutual dependence formally 
should correspond to specific structure of the world, in which unified treatment of internal degrees 
of freedom (reduced to gauge symmetries) and space-time coordinates may not be completely 
covered by the string theory. Some more general mathematical framework may be appropriate, e.g. 
the invariant formulation of the apparatus of fiber bundle theory (see \cite{3}  and Refs. wherein), 
treating space-time and internal variables on equal footing and making it possible to fix the true 
symmetry of fundamental interactions and geometry of space-time from exact solutions for the 
functional integral. The realization of such program can lead to the true physically selfconsistent 
theory of space-time, elementary particles and fundamental natural forces. As a step in this 
direction, elaboration of unified models of cosmology and particle physics is important.
Such models treat physically selfconsistent complete cosmological scenarios. Physical 
selfconsitency means, that the physical grounds for inflation, baryosynthesis and dark matter are 
considered in the unified theoretical framework on the base of the unique particle model, and the 
degree of completeness assumes the accuracy, with which the astronomical observational data are 
reproduced in the considered cosmological scenario. The degree of completeness of the 
cosmological model should depend on the properties of the physical model only.

The easiest way to construct cosmologically selfconsistent particle models is to extend the SM by 
addition to its $SU(3)_C\bigotimes SU(2)_W\bigotimes U(1)_Y $
 symmetry some other global or local gauge symmetries or by inclusion of the SM symmetry group 
into more general gauge group. As a result, the extended gauge model contains new particles and 
fields, related to new symmetries added to the standard model. In the most cases, the masses of new 
particles and strength of new interactions, mediated by new fields, correspond to superhigh energy 
scales, inaccessible to direct experimental test at accelerators.  At best, experimental high-energy 
physics can put lower limits on some parameters, related to these scales. The only possibility is to 
elaborate a system of indirect physical, astrophysical and cosmological constraints on the free 
parameters of the "hidden" sector of particle model, to fix them and to specify the cosmological 
scenario, following from this choice.

The strategy of cosmoparticle physics approach to unified models of cosmology and particles can 
be stipulated as follows:
\begin{enumerate}
\item Physically motivated choice for extended gauge particle model.
\item Test for its cosmological selfconsistency -- study of its possibility to reproduce cosmological 
and astrophysical phenomena and effects
\item Determination of free parameters of the "hidden" sector of particle model or set of constraints 
on them from the combination of indirect cosmological, astrophysical and experimental physical 
restrictions.
\item Elaboration of complete quantitatively definite cosmological scenario.
\item Formulation of the system of indirect experimental physical and astronomical effects, 
providing the detailed test of the physical model and cosmological scenario, based on it.
\item Estimation of completeness of this scenario.
\end{enumerate}
 
Cosmoparticle physics puts traditional methods of observational astronomy and experimental 
physics into nontrivial multidimensional complex system of links, thus enriching substantially the 
collaboration between physics and astronomy established by astroparticle physics.

\section{The system of links between astronomical observations and laboratory physics 
experiments}
Links between particle physics and cosmology are generally viewed by astroparticle physics as 
system of linear relations. So, statements \cite{4}, that  electron neutrino mass is about  $30eV$, 
immediately lead to cosmological consequences, since Big Bang cosmology predicts primordial 
neutrino background with the concentration, equal to $3/11$ of the one of relic photons. By 
multiplying the neutrino mass on the concentration of cosmological neutrino background one 
immediately found, that the massive neutrino density should dominate in the modern Universe and 
that gravitational instability in the nonrelativistic gas of massive neutrinos should play the dominant 
role in the formation of the large scale structure of the Universe. Primordial massive neutrinos were 
identified with the hot dark matter in the halo being one of the three classes of elementary particle 
dark matter (DM) candidates.

In general hot DM refers to low mass neutral particles that where still in thermal equilibrium  after 
the QCD phase transition. Hot DM particles have a cosmological number density roughly 
comparable to that  of microwave background photons, which implies an upper limit to their mass 
of a few ten $eV$. Neutrinos are the standard example of hot DM, although other possibilities such 
as Majorons are discussed in the literature. Majorons are the pseudo-Goldstone bosons connected 
with the Majorana nature of the mass of neutrino. Majorana mass of neutrino corresponds to lepton 
number violation. In this case lepton number violating processes such as nuclear neutrinoless 
double beta decay can take place. If at least two types of neutrino are massive and neutrino states of 
definite mass do not coincide with the states with definite lepton number, neutrino oscillations 
should take place. In the matter resonant enhancement of neutrino oscillations can take place, what 
may be the solution for Solar neutrino puzzle at very small values of the difference of neutrino mass 
squares $\delta m^2\simeq10^{-6}eV$. The detailed analysis of all these crossdisciplinary links, 
undertaken by astroparticle physics, could not however lead to any definite conclusion in view of 
evident troubles of the simple model of massive electron neutrinos in its confrontation with the 
observational and experimental data.

	The successive experimental measurements of electron neutrino mass in studies of beta 
spectrum of tritium lead to ambiguous results, not confirming the original claims on the value of 
$\simeq 30eV$. The upper limit on the electron neutrino mass is roughly $10eV\div 15eV$, a more 
precise limit cannot be given since unexplained effects have resulted in the negative value of 
$m(\nu_e)^2$ in recent tritium beta decay experiments. The (90\% C.L.) upper limit on an effective 
Majorana neutrino mass $0.65eV$ from Heidelberg-Moscow $^{76}Ge$ neutrinoless $2\beta$ 
decay experiments \cite{5}. The upper limits from accelerator experiments on the masses of the 
other neutrinos are $m(\nu_{\mu})<0.17MeV$ and $m(\nu_{\tau})<24MeV$ (95\% C.L.). The 
events that appear to represent $\bar\nu_{\mu}\to\bar\nu_e$ oscillations followed by 
$\bar\nu_e+p\to n+e^+$, $n+p\to D+\gamma$, with coincident detection of $e^+$ and the 
$2.2MeV$ neutron-capture $\gamma$ ray in the Liquid Scintillator Neutrino Detector (LSND) 
experiment at Los Alamos suggest that 
$\Delta m^2_{e\mu }=\mid m(\nu_{\mu})^2- m(\nu_{e})^2\mid >0$   \cite{6}. Comparison with 
exclusion plots from other experiments implies a lover limit 
$\Delta m^2_{e\mu }=\mid m(\nu_{\mu})^2- m(\nu_{e})^2\mid >0.2eV^2$, implying in turn a 
lower limit $m_{\nu}\ge 0.45eV$, or $\Omega_{\nu}\ge0.02(0.5/h)^2$. More data and analysis are 
needed from LSND's $\nu_{\mu}\to\nu_e$ channel before the initial hint \cite{7} that 
$\Delta m^2_{\mu e}\approx 6eV^2$ can be confirmed. Recent Super-Kamiokande data following 
the Kamiokande data \cite{8} show that the deficit of $E>1.3GeV$ atmospheric $\nu_{\mu}$ 
increases with zenith angle. These data suggested that $\nu_{\mu}\to\nu_{\tau}$ oscillations length 
is comparable to the height of the atmosphere, implying that  
$\Delta m^2_{\tau\mu}\simeq 10^{-3}eV^2$ -- which in turn implies that if either $\nu_{\mu}$
or $\nu_{\tau}$ have large enough mass ($\ge 1eV$) to be a hot dark matter particles, then they 
must be nearly degenerate in mass, i.e., the hot dark matter mass is shared between these two 
neutrino species. However, the deficit of atmospheric $\nu_{\mu}$ even at small zenith angles, 
corresponding to paths much smaller than oscillation length, causes serious doubts in the 
interpretation of Super-Kamiokande and Kamiokande data \cite{9}. At 
$\Omega_{\nu}\simeq 1$ neutrino free streaming strongly suppresses adiabatic fluctuations at 
scales smaller than galaxy superclusters ($\simeq 10^{15}M_{\odot}$). With the use of the COBE 
upper limit, hot DM with adiabatic fluctuations would hardly lead to any structure formation at all. 
The proper choice of a possible solution for this problem - transition to more complicated cases, hot 
DM plus some sort of seeds, such as cosmic strings (see for example \cite{10}) or to other class of 
dark matter candidates, corresponding to cold DM (CDM) scenario - has in fact no fundamental 
grounds in the framework of astroparticle physics. Moreover the physical grounds for neutrino 
instability or for CDM particles are not alternative to the ones for neutrino rest mass, and from the 
physical viewpoint the general case should account for all these possibilities. Cold DM consists of 
particles for which the scale of free streaming is very small and its existence leads to strong  
dynamical effects at galaxy scale.

The development of CDM models and their troubles in the framework of astroparticle physics seem 
to confirm the general wisdom on true complexity of the world system. The two sorts for cold DM 
that are best motivated remain supersymmetric particles (WIMPs)  and axions.

Supesymmetry  underlies almost all new ideas in particle physics, including superstrings. There are 
two key feature of supesymmetry that make it especially relevant to DM, $R$ -- parity and the 
connection between supersymmetry breaking and the electroweak scale. The $R$ -- parity of any 
particle is $R\equiv (-1)^{L+B+S}$, where $L$, $B$, and  $S$ are its lepton number, baryon 
number, and spin. In most version of supersymmetry, $R$ -- parity is exactly conserved. This has 
the powerful consequence that the lightest $R$ -- odd particle -- often called the "lightest 
supersymmetric partner" (LSP)- must be stable, for there is no lighter $R$ -- odd particle for it to 
decay into. The LSP is thus natural candidate to be the dark matter. In the standard version of 
supersymmetry, there is an answer to the deep puzzle why there should be such a large difference in 
mass between the GUT scale $M_{GUT}\simeq 10^{16}GeV$ and the electroweak scale 
$M_W=80GeV$. Since both gauge symmetries are supposed to be broken by Higgs bosons which 
moreover must interact with each other, the natural expectation would be that 
$M_{GUT}\simeq M_W$
or that $M_W$  is induced by radiative correction $M_W\sim\alpha M_{GUT}$. The 
supersymmetric answer to this "gauge hierarchy" problem is that the masses of the weak boson 
$W^{\pm}$ and all other light particles are zero until supersymmetry itself breaks. Thus, there is a 
close relationship between  the masses of the supersymmetric partner particles and the electroweak 
scale. Since the abundance of the LSP is determined by its annihilation in the early Universe, and 
the corresponding cross section involves exchanges of weak bosons or supersymmetric particles -- all 
of which have electromagnetic-strength couplings and masses $\simeq M_W$ -- the cross section 
will be $\sigma\simeq e^2s/M^4_W$ (where $s$ is the square of the center of mass energy) i.e., 
comparable to the that of typical weak interaction processes. This in turn has the remarkable  
consequence that the modern density of LSPs can be close to the critical density, i.e. 
$\Omega_{LSP}\simeq 1$.  The LSP is in the most cases a spin -- $1/2$ Majorana particle called 
"neutralino", which represents the linear combination of photino (supersymmetric partner of the 
photon), zino (partner of the $Z^0$), Higgsinos (partners of the two Higgs bosons associated with 
electroweak symmetry breaking in supersymmetric theory), and axinos (partner of the axion). 
Neutralinos are Weakly Interacting Massive Particles (WIMPs) with the mass from tens to hundreds 
GeV, and thus are natural candidates for the cold DM.

 The prediction of invisible axion follows from another line of theoretical argumentation, related to 
the solution of the strong CP violation problem in QCD. Searches for axion emission in $\mu$,  
$K$ decays and nuclear decays put lower limit on the scale of axion physics. Constraints on stellar 
energy losses due to axion emission put this limit even higher: up to $10^6GeV$ in the case of 
archion and up to $10^8GeV$ for the bulk of other invisible axion models. In cosmology, 
primordial coherent axion field oscillations were found to behave in respect to gravitational 
instability as gas of very heavy particles, making invisible axion popular CDM candidate. 
Experimental searches for cosmic and Solar axion fluxes are under way, based on the predicted 
effect of axion-photon conversion in time--varying electromagnetic field.

In the framework of astroparticle physics it is not possible to find physical motivations which 
candidate on CDM particle -- neutralino or axion -- is more preferable. From particle physics 
viewpoint the both candidates are important, since both supersymmetry and invisible axion solution 
are necessary to remove internal inconsistencies of the standard model: supersymmetry removes 
quadratic divergence of Higgs boson mass in the electroweak theory and axion recovers from strong 
CP violation in QCD. Astroparticle physics has no theoretical tools to find the proper combination 
for the both hypothetical phenomena. Moreover, recent analysis of the observational data on the 
large scale structure and of the anisotropy of thermal electromagnetic background find troubles in 
simple CDM model and favors more sophisticated dark matter scenario, such as mixed cold+hot 
dark matter (see for example \cite{11}). It appeals for necessity in special methods to deal with 
multiparameter space of physical and cosmological parameters, which astroparticle physics does not 
possess.

Together with the proper combination of studies of cosmological large scale structure, relic 
radiation, nucleosynthesis, tests for inflational, baryosynthesis and dark matter models 
cosmoparticle physics invokes such forms of crossdisciplinary studies as cosmoarcheology or 
experimental physical cosmology.

\section{Cosmoparticle approach to the problem of fermion masses and mixing}
The problem of fermion families is one of key problems in the modern particle physics. It has 
different aspects, questioning the origin of family replication, quark and lepton mass spectrum and 
mixing pattern, CP violation in weak interactions, CP conservation in strong interactions, 
suppression of flavor changing neutral currents (FCNC), pattern of neutrino masses and oscillations, 
etc. Thus the particle model of fermion families should offer the solution to all these problems. 
The standard model (SM) is successful in describing various experimental data (see for example 
\cite{12}) and it can be considered as a minimal necessary element of any theory of flavor. In SM 
the three families, sharing the same quantum numbers under the 
$SU(3)_C\bigotimes SU(2)_W\bigotimes U(1)_Y $ gauge symmetry, are introduced as an anomaly 
free set of chiral left-handed fermions $q_i=(u_i,d_i)$, $u^c_i$, $d^c_i$; 
$l_i(\nu_i,e_i)$, $e^c_i$, 
where $i=1,2,3$ is a family index. In SM the masses of fermions and $W^{\pm}$, $Z$ gauge 
bosons have the common origin in the Higgs mechanism. Quarks and charged leptons get masses 
through the Yukawa couplings to the Higgs doublet $\phi$:
\beq
\label{1}
L_{Yuk}=\lambda^u_{ij}q_iCu^c_j\tilde\phi +\lambda^d_{ij}q_iCd^c_j\phi +
\lambda^e_{ij}l_iCe^c_j\phi\qquad (\tilde\phi =i\tau_2\phi^*)
\eeq
So, the fermion masses are related to the weak scale 
$\langle\phi\rangle =v=174GeV$. However, the Yukawa constants are arbitrary, namely 
$\hat\lambda^{u,d,e}$ are in general complex $3\times 3$ matrices. To reproduce the masses of 
quarks and leptons one has to put by hands $27$ values of these matrix elements. The SM contains 
no renormalizable couplings that could generate the neutrino masses:
\beq
\label{2}
L_{\nu}=\frac{\lambda^{\nu}_{ij}}{M}(l_i\tilde\phi )C(l_j\tilde\phi ),\qquad
\lambda^{\nu}_{ij}=\lambda^{\nu}_{ji}
\eeq
where $M>>v$ is the regulator mass, which depends on the mechanism of neutrino mass ({2}) 
generation. The matrices of coupling constants and the corresponding fermion mass matrices 
$\hat m^f=\hat\lambda^fv$ $(f=u,d,e)$ and $\hat m^{\nu}=\hat\lambda^{\nu}(\nu^2/M)$
can be reduced to the diagonal form by the unitary transformations $V_f$ and $V_{\nu}$. Hence, 
quarks are mixed in the charged current interactions, and these mixings are determined by Cabibbo-
Kobayashi-Maskawa (CKM) matrix. The CKM matrix is parameterized by three mixing angles and 
CP-violating phase. In the case of massive neutrinos, a similar mixing matrix emerges also in the 
lepton sector. 
The fermion family puzzle consists in the following phenomena: 
the mass spectrum of quarks and charged leptons is spread over five orders of magnitude, from 
MeVs to hundred GeVs;
the weak transitions dominantly occur inside the families, and are suppressed between different 
families thereby the SM exhibits the natural suppression of the flavor changing neutral currents 
(FCNC), both in  the gauge boson and Higgs exchanges;
the Yukawa constants in {1} are generally complex, the observed CP- violating phenomena can be 
explained by the CKM mechanism with sufficiently large CP-phase $\simeq 1$. However, at the 
same time it induces the strong CP violation problem (see for example \cite{13}): the overall phase 
of the Yukawa matrices gives effective contribution to the vacuum $\Theta$- term in QCD and thus 
induces the P and CP violation in strong interactions. On the other hand, the measurements of 
dipole electric moment of neutron impose  the strong bound  $\Theta <10^{-9}$;
the experimental data show some ambiguous indications for neutrino masses and mixing.
 The fermion mass and mixing problem can be formulated as a problem of matrices of the Yukawa 
couplings $\hat\lambda^f$, which remain arbitrary in the SM. There is no explanation, what is the 
origin of the observed hierarchy between their eigenvalues, why $\hat\lambda^u$ and 
$\hat\lambda^d$ are small, what is the origin of the complex structure needed for the CP- violation 
in weak interactions, why the $\Theta$- term is vanishingly small in spite of the complex Yukawa 
matrices. It is attractive to think that at some scale above the electroweak scale there exists a more 
fundamental theory which could allow to calculate the Yukawa couplings, or at least to fix the 
relationship between them.
  
The structure of mass matrix can be related with the spontaneously broken horizontal symmetry 
between fermion families. Consider, for example, model with all quark and lepton states 
transforming as triplets 
$f_{\alpha}=(q,l,u^c,d^c,e^c)_{\alpha}$ of the horizontal $SU(3)_H$ symmetry \cite{14},  
($\alpha =1,2,3$ is a family index). Such a horizontal symmetry does not allow quarks and leptons 
to have renormalizable Yukawa couplings. Thus, the fermion mass generation is possible only after 
the $SU(3)_H$ breaking, through the high order (non-renormalizable) operators (HOPs) involving 
some "horizontal" Higgses inducing this breaking at the scale $V_H>>v$. This suggests that the 
observed mass hierarchy may emerge due to the hierarchy in the $SU(3)_H$ breaking.  Full 
$SU(3)_H$ breaking is achieved by introducing the horizontal scalars: a sextet 
$\chi_3^{\{\alpha\beta\}}$
 and two other sextets or triplets 
$\chi_{1,2}^{[\alpha\beta ]}\simeq\varepsilon^{\alpha\beta\gamma}\chi_{\gamma}$. 
The pattern of their $3\times 3$ VEV matrix can be chosen so that the first sextet VEV is acquired 
by $(3,3)$ component, and in sextets (or triplets)  $\chi_2$ and $\chi_1$ the smaller VEVs 
$V_{23}$ and $V_{12}$ are acquired by $(2,3)$ and $(1,2)$ (or first and third ) components. 
VEVs follow the hierarchy $V_{33}>>V_{23}>>V_{12}$, which is stable relative to radiative 
corrections. Thus in the context of the $SU(5)\otimes SU(3)_H$ theory with fermions in 
representations $(\bar 5+10)_{\alpha}$, the relevant HOPs \cite{14} can be induced through the 
renormalizable interactions, as a result of integrating out the effects of hypothetical superheavy 
particles (see, for example, \cite{15,16}). In the other words, the quark and lepton masses can be 
induced through their mixing with superheavy  -- fermions, in a direct analogy to the see--saw 
mechanism of neutrino mass generation.  In this case the VEV pattern of Higgs multiplets $\chi$ is 
reflected in the Yukawa matrices, and the fermion mass hierarchy follows the hierarchy of 
$SU(3)_H$ symmetry breaking. There are two possible choices for the representation of  $F$ -- 
fermions, and, respectively, one can generate two types of the pattern of Yukawa mass matrices 
\cite{17,18}.  The first case corresponds to a direct hierarchy pattern. In particular, the VEV pattern 
leads directly to the Fritzch texture. Another possibility is the inverse hierarchy. In the latter case 
the VEV pattern is inverted in the fermion mass structure (see more detail \cite{17,18,16}).
Thus, the horizontal $SU(3)_H$ symmetry is attractive since it unifies all families. For the solution 
of the strong CP- problem one can introduce the Peccei-Quinn (PQ) type symmetries \cite{19}, 
which in additionally could further restrict the mass matrix structure. In particular, in the horizontal 
$SU(3)_H$  symmetry models the PQ symmetry can be naturally related to the phase 
transformation of the horizontal scalars $\chi$ \cite{17,18}. 
Consider as an example the application of the approach of cosmoparticle physics (section 2) to the 
problem of fermion flavours. This strategy can be stipulated as follows.

{\bf Step 1.}  The class of physically motivated extensions of SM is considered, namely, the class 
of gauge models with horizontal  family symmetry.

{\bf Step 2.} The inevitable consequences of chosen class of models, which are able to reproduce 
cosmological and astrophysical phenomena and effects are the following:
\begin{itemize}
\item the existence of the specific type of invisible axion (archion), which is simultaneously 
Majoron and familon \cite{17,18};
\item the existence of horizontal scalars $\chi$ with superhigh energy scale of VEVs;
\item the existence of neutrino Majorana mass with the hierarchy of neutrino masses;
\item the nonconservation of lepton number $\Delta L=2$;
\item the instability of neutrino relative to decays on more light neutrino and archion;
\item the Dirac see-saw mechanism and singlet scalar $\eta$ , connected with it;  
\end{itemize}

{\bf Step 3.}  One introduces the main free parameter $V_H$ of the hidden sector of the considered 
model, namely, the scale of horizontal $SU(3)_H$ symmetry breaking. The set of indirect 
cosmological, astrophysical and experimental physical restrictions on the hidden sector is revealed 
from following phenomena:
\begin{itemize}
\item from the analysis of data of nondiagonal transitions  (for example $\mu\to ea$ and $K\to\pi a$ 
(where $a$ is archion)) \cite{20,21};
\item from the astrophysical estimations of  stellar energy losses due to archion emission \cite{18};
\item from the analysis of archion emission influence the time scale and  energetics of neutrino flux 
from collapsing star \cite{18};
\item from the analysis of inhomogeneities generated by the large scale modulation of coherent 
axion field oscillations \cite {22,23,24};
\item from the analysis of primordial black holes formation in the second order phase transitions 
connected with three stages of horizontal $SU(3)_H$ -- symmetry breaking, which take place at the 
inflationary stage \cite{25,24};
\item from the effect of nonthermal horizontal symmetry restoration at postinflational dust-like 
stage \cite{24,26};
\end{itemize}
Taking together all limits imposed by the pointed phenomena it is possible to extract two narrow 
windows for the value of the parameter $V_H$. They are the "low" energy branch $V_6$ 
\cite{23,27} and the "high" energy branch $V_{10}$ \cite{24}.

{\bf Step 4.}  With the use of the above restrictions one can elaborate the physically motivated full 
cosmological model, which is based on the chosen horizontal extension of SM. This model has been 
called the model of "horizontal" unification (MHU) \cite{23,24}.
\begin{itemize}
\item MHU solves the problems of SM connected with family problem and strong CP violation 
problem in QCD; it predicts qualitatively new type of invisible axion (archion) \cite{17,18,28}; it 
predicts the neutrino masses and neutrino flavour nondiagonal transitions with emission of archion.
\item MHU predicts the following history of Universe: 
\begin{itemize}
\item The early Universe starts from the inflational stage \cite{23,24}, driven by the inflaton field 
$\eta$, being singlet relative to all gauge groups. The VEV of this field plays the role of the 
universal energy scale in the Dirac see-saw mechanism of the generation of masses of charged 
fermions \cite{16,17,23,24}. When the inflational stage is finished the inflaton field decays due to 
interactions assumed by the Dirac see-saw mechanism \cite{23,24}. It leads to reheating of the 
Universe and consequently to transition to the standard Friedman cosmology.
\item The reheating temperature is sufficiently high for generation of the observed baryon 
asymmetry. The baryogenesis mechanism in the MHU combines the $(B+L)$ nonperturbative 
electroweak nonconservation at high temperatures with $\Delta L=2$ nonequilibrium transitions, 
induced by Majorana neutrino interaction \cite{23}. The mechanism can provide inhomogeneous 
baryosynthesis and even to the existence of antimatter domains in baryon asymmetrical Universe 
\cite{24}.
\item There are two possible scenarios of large scale structure (LSS) formation:
\begin{itemize}
\item Hierarchic decay scenario (HDS) \cite{23,21}, realized at the "low" energetic scale ($V_6$). 
In the HDS the LSS formation takes place in the succession of stages of dominance of unstable 
neutrino and their relativistic decay products.
\item Mixed stable dark matter, realized at "high" energetic scale ($V_{10}$) \cite{24}. The 
formation of LSS in this case occurs at the conditions of dominance of coherent oscillations of 
axion field and massive stable neutrino (see \cite{24} in more detail).
\end{itemize}
\end{itemize}
\end{itemize}

{\bf Step 5.}  The system of the detailed indirect test of MHU and MHU-based cosmological 
scenario can use the following signatures:
\begin{itemize}
\item MHU predicts flavour nondiagonal decays of leptons, mesons and hyperons (see \cite{21,27} 
in more detail);
\item MHU predicts the level of oscillations $K\to\bar K$,  $B\to\bar B$ \cite{21};
\item astronomical search for invisible axions (see for example \cite{29}) and their two-photon 
decays;
\item experimental searches for solar axions (see for example \cite{30});
\item experimental searches for the force, violating the Equivalence Principle, which is connected 
with the existence of invisible axion (see for example \cite{31}).
\end{itemize}

{\bf Step 6.}  The estimation of completeness of obtained scenario is necessary to determine the 
direction of the further extension of the considered approach. In the other words the elaborated 
cosmological model should incorporate the cosmological consequence of some other extensions of 
the SM such as GUT and SUSY. In particular, the estimation of completeness of MHU can be 
obtained by the comparison of the predicted consequences of the MHU-based scenario of inflation, 
baryosynthesis and LSS formation with the astronomical observations (see \cite{24} in more 
details).

To conclude, the development of cosmology and particle physics and the nontrivial tests of their 
foundations in combination of indirect evidences follow the laws of cosmoparticle physics, that will 
unify on the basis of its principles the existing trends in studies of mutual relationship of elementary 
particles and the Universe, widely represented in the present proceedings.

\end{document}